\documentclass[]{spie}  

\usepackage{amsmath,amsfonts,amssymb}
\usepackage{graphicx}
\usepackage[colorlinks=true, allcolors=blue]{hyperref}
\usepackage{xcolor}

\usepackage{listings}
\title{The future looks dark: improving high contrast imaging with hyper-parameter optimization for data-driven predictive wavefront control}

\author[a]{J. Fowler}
\author[a]{Rebecca Jensen-Clem}
\author[b]{Maaike A. M. van Kooten}
\author[a]{Vincent Chambouleryon}
\author[c]{Sylvain Cetre}
\affil[a]{Department of Astronomy \& Astrophysics, University of California, Santa Cruz, CA, USA}
\affil[b]{National Research Council Canada, Herzberg Astronomy and Astrophysics Research Center, Victoria, Canada}
\affil[c]{W. M. Keck Observatory, Waimea, Hawaii, USA}

\authorinfo{Further author information: (Send correspondence to J. F.)\\J. F.: E-mail: jumfowle@ucsc.edu}

\begin{document} 
\maketitle

\begin{abstract}
The direct imaging and characterization of exoplanets requires extreme adaptive optics (XAO), achieving exquisite wavefront correction (upwards of 90$\%$ Strehl) over a narrow field of view (a few arcseconds). For these XAO systems the temporal error is often a leading term in the error budget, wherein the wavefront evolves faster than the lag between wavefront sensing and control. For atmospheres with high-velocity wind layers, this can result in a wind-driven halo in the coronagraphic dark-zone, limiting sensitivity to faint, close-in companions. The AO system's lag-time is often limited by the wavefront sensor exposure time, especially in the case of fainter guidestars. Predictive control mitigates the temporal error by predicting the shape of the wavefront by time the system correction is applied. One such method of prediction is empirical orthogonal functions (EOF), wherein previous states in the wavefront sensor history are used to learn linear correlations with a minimization problem. This method has been demonstrated on-sky at Subaru/SCExAO and Keck/NIRC2, but has yet to be optimized. With this work as a starting point, we explore the optimal filter hyper-parameter space for implementing EOF on-sky, study its stability under varying atmospheric parameters, and discuss future paths for facilitization of predictive control. This work not only offers a pathway to optimize Keck and Subaru observing, but also acts as a pathfinder for predictive control methods with extremely large telescopes.
\end{abstract}

\keywords{predictive control, temporal error, empirical orthogonal functions (EOF), hyper-parameter optimization, simulated annealing, Keck, Subaru}

\section{INTRODUCTION}
\label{sec:intro}  

Predictive wavefront control, first suggested by Dessenne in 1997 \cite{Dessenne1997}, has been explored through many implementations over the years, ranging from classic and predictive Linear Quadratic Gaussian controllers \cite{Sivo2014, Poyneer2023}, neural network approaches \cite{Wong2021, Swanson2021}, through reinforcement learning \cite{Haffert2021, Landman2021, Nousiainen2024, Pou2022}, and through classic minimization problems \cite{Guyon2017, Jensen-Clem2019, vanKooten2020, Fowler2022}; for a more complete review of predictive methods, see (Fowler, 2023)\cite{Fowler2023}. Predictive controllers consistently show promising performance with on-sky testing \cite{Guyon2020, Dessenne1998, Marquis2024}, and in particular, empirical orthogonal functions (EOF) at Keck has shown up to 3x improvement in contrast at separations of 3-7 $\lambda$/D, and 2x for 3 $\lambda$/D\cite{vanKooten2022}. 

However, no facilitized implementations of predictive control are running on-sky and little has been published in the way of optimizing its performance for regular usage under varying atmospheric conditions. The implementation at Keck\cite{vanKooten2022} explored the optimal mixing between an integrator and predictor and optimal gain for the high order correction, but built a predictive filter with a consistent set of hyper-parameters. Similarly, one work\cite{Wong2021} explored the optimal amount of data for an EOF filter to train on with Subaru/SCExAO telemetry, but predictive control at SCExAO is run on-sky with a static set of filter hyper-parameters (private communication, SCExAO team). 

In the effort to move predictive control from an on-sky demonstration to a facilitized mode, we must address how to consistently set up the predictive controller for varying nightly conditions and make recommendations for a hyper-parameter optimization strategy during observing nights. Here, we address the tuning of individual hyper-parameters in the predictive filter as the first step towards developing such an on-sky strategy.

\section{TUNEABLE HYPER-PARAMETERS IN THE EOF FILTER}

\begin{figure}[h!]
\makebox[\textwidth][c]{
    \mbox{\includegraphics[width=0.7\textwidth]{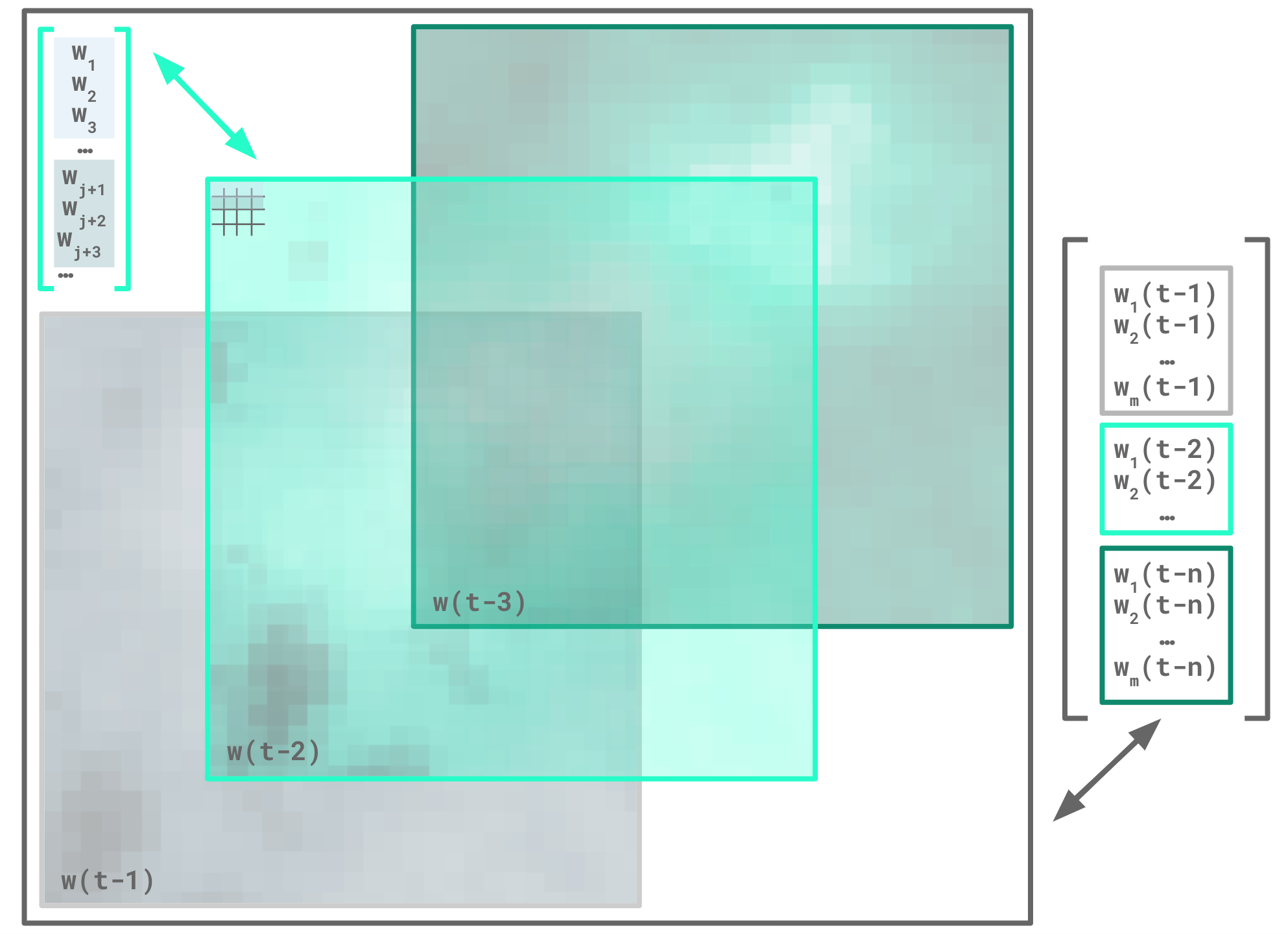}}
    }
    \caption{The history vector contains \textbf{n} frames of wavefront sensor data flattened into a single vector.}
\label{fig:history_vector}
\end{figure}

Empirical orthogonal functions (EOF)\cite{Guyon2017} builds a linear filter \textbf{F} by assembling \textbf{l} previous states of wavefront sensor information (or history vectors \textbf{h}) into a training matrix \textbf{D}, and comparing those to known future states \textbf{P}. The history vector contains \textbf{n} frames of wavefront sensor data (each with \textbf{m} points for each mode) flattened into a single vector; this is shown visually in Figure \ref{fig:history_vector}.
The filter matrix encodes the wind velocities and other linear system errors. 

The logic to build a filter matrix is shown below; the matrix inversion is done with a least-squares inverse, and uses $\alpha$ as a regularization parameter.
\begin{align}
    &min||\mathbf{D}^T\mathbf{F}^T - \mathbf{P}^T||^2 \\
     &\mathbf{F} = ((\mathbf{D}^T)^\dagger \mathbf{P}^T)^T  \\
     &\mathbf{F} = \mathbf{P}\mathbf{D}^T(\mathbf{D}\mathbf{D}^T + \alpha\mathbf{I})^{-1}  \\
     &\textrm{prediction} = \mathbf{F}\mathbf{h}
\end{align}

\begin{figure}[h!]
\makebox[\textwidth][c]{
    \mbox{\includegraphics[width=0.7\textwidth]{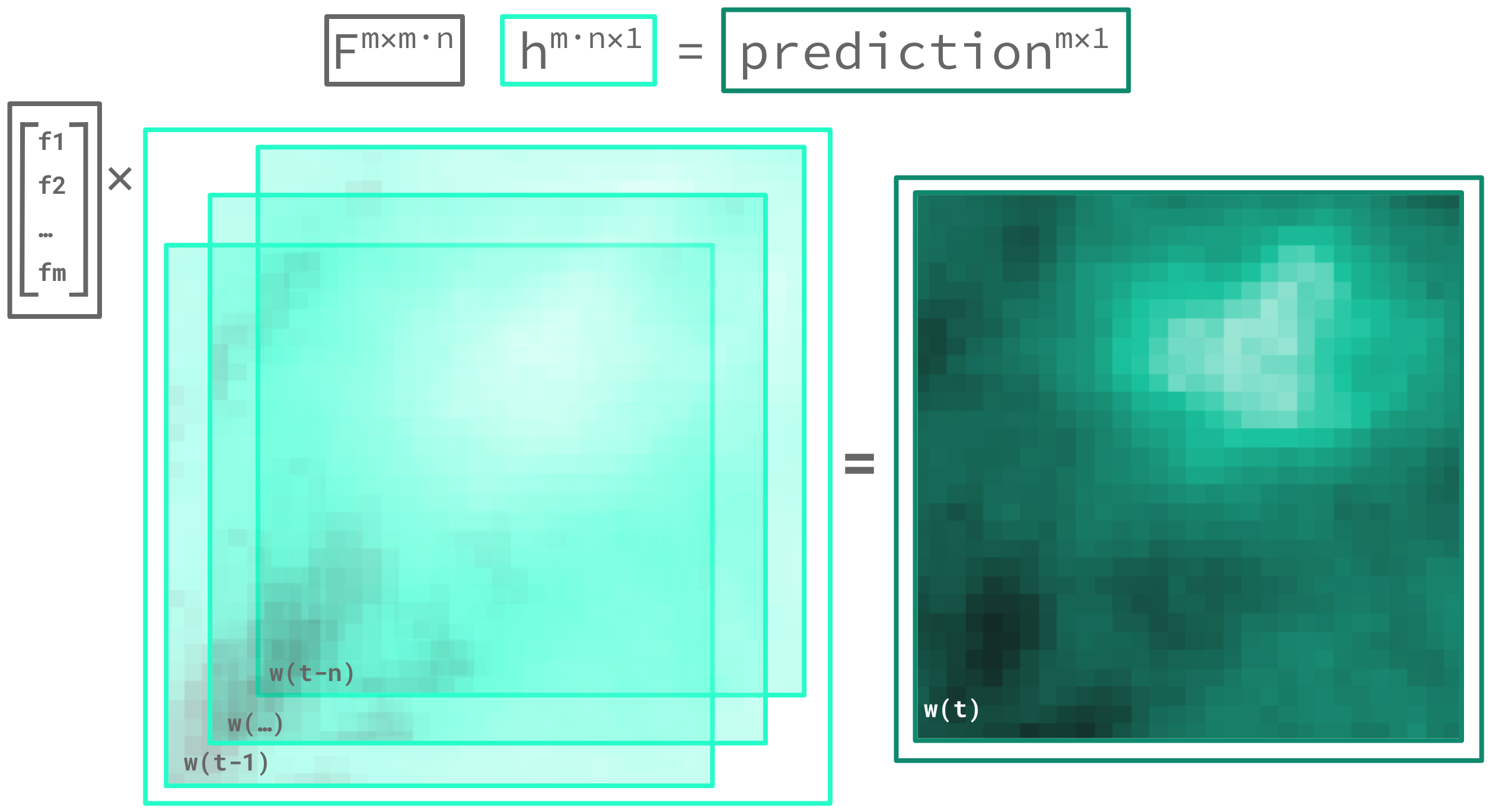}}
    }
    \caption{To predict a future iteration with \textbf{m} modes, we multiply a filter matrix that is \textbf{m} by \textbf{m}$\cdot$\textbf{n} (with \textbf{m} modes and \textbf{n} points in time) by the current history vector \textbf{h}. }
\label{fig:pwfc_diagram}
\end{figure}

At every iteration of the real-time-controller, we multiply the filter matrix \textbf{F}, by the history vector \textbf{h} to predict an iteration one lag-time in the future. This full logic is visualized in Figure \ref{fig:pwfc_diagram}. 
 
The filter matrix provides a good correction while the atmosphere is driven by the same linear forces on which it was trained, therefore we must plan to update the filter matrix every few minutes or have a rolling update of the filter matrix for on-sky operations. To generate a predictive filter, we consider the number of frames in our training data \textbf{l}, the number of frames in our history vector \textbf{n}, and the regularization constant $\alpha$ as tuneable hyper-parameters of the EOF filter. In simulation, our matrix is well conditioned, and we can run with an $\alpha=1$, so the simulations presented in the rest of this work aim to explore optimal values for \textbf{l} and \textbf{n}.

\section{OPTIMIZING FILTER HYPER-PARAMETERS IN SIMULATION}
\label{sec:sections}

We simulate the performance of EOF with varying hyper-parameters \textbf{l} and \textbf{n} as applied to atmospheres with varying windspeeds and Fried parameters. We simulate a classic single-layer Kolmogorov\cite{Kolmogorov1941} phase screen with the Taylor frozen flow approximation\cite{Taylor1938} with a given Fried parameter r$_0$ and windspeed for a 10 meter telescope at a resolution of 48x48 pixels. We subtract piston, tip, and tilt modes, and apply a circular aperture. We sample and correct the atmosphere at 1kHz, with a lag of two time steps (2 ms), with perfect wavefront sensing and perfect correction -- so the error is indicative only of error from the control itself. 

\begin{figure} [ht]
   \begin{center}
   \begin{tabular}{c} 
   \includegraphics[width=0.32\linewidth]{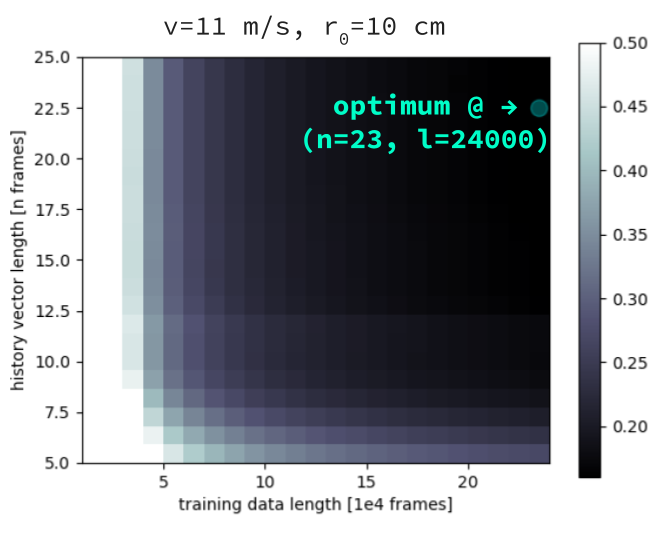}
\includegraphics[width=0.32\linewidth]{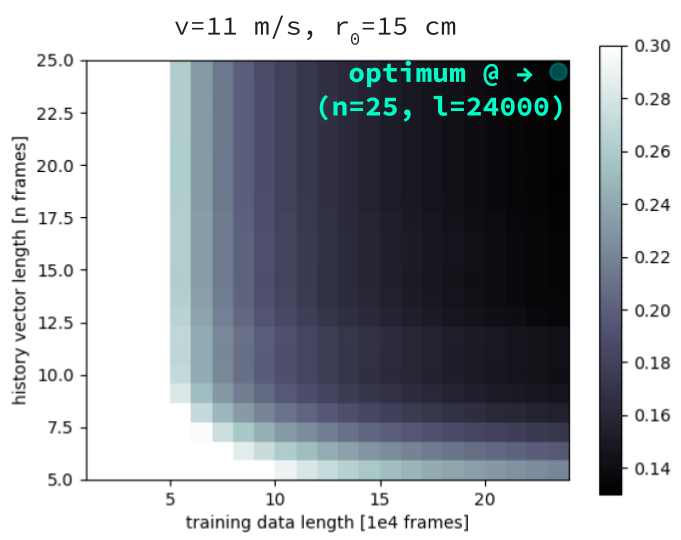}
\includegraphics[width=0.32\linewidth]{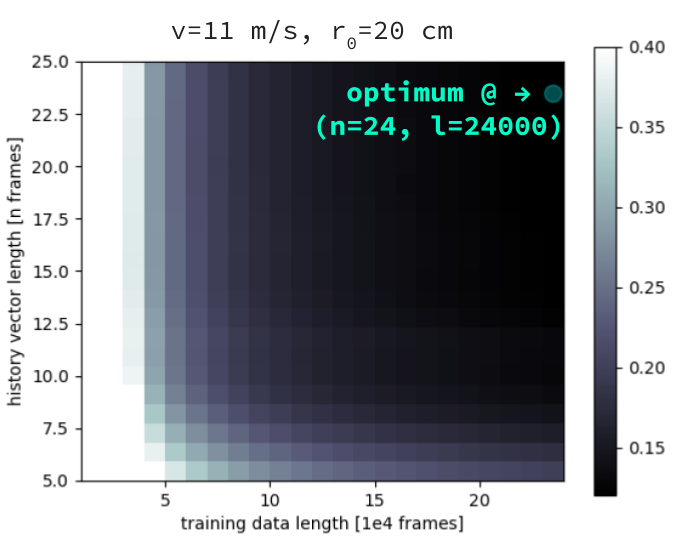}
   \end{tabular}
   \end{center}
   \caption[example] 
   { \label{fig:same_speed} Heatmap of EOF predictor performance (in median RMS error in nm) as training data length \textbf{l} and history vector length \textbf{n} vary. Left: r$_0$ of 10 cm. Middle: r$_0$ of 15 cm. Right: r$_0$ of 20 cm. All of the above simulations have a windspeed of 11 m/s. Note that the optimal history vector length changes based on r$_0$.}
   \end{figure}

\begin{figure} [ht]
   \begin{center}
   \begin{tabular}{c} 
   \includegraphics[width=0.32\linewidth]{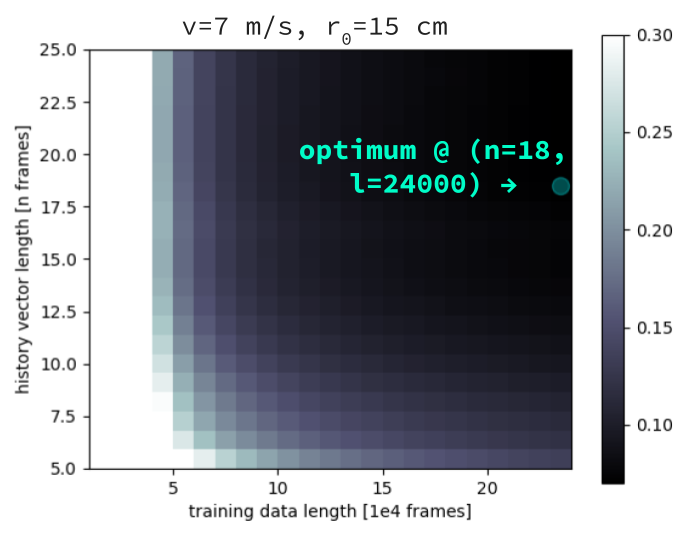}
\includegraphics[width=0.32\linewidth]{figures/v11_r015_label.png}
\includegraphics[width=0.32\linewidth]{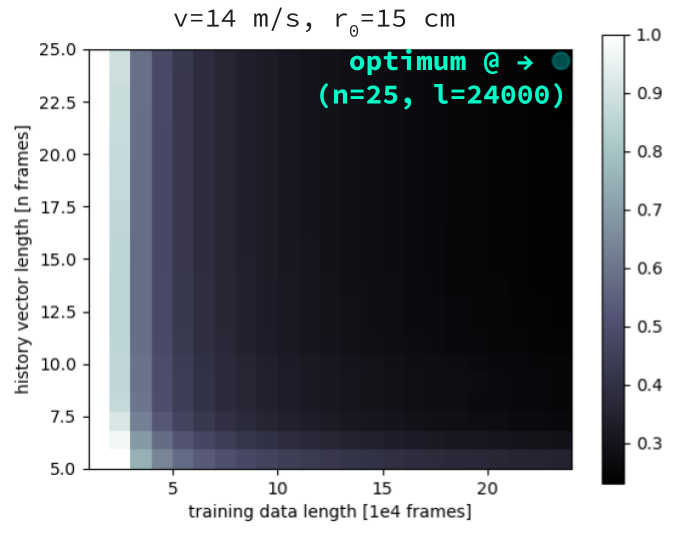}
   \end{tabular}
   \end{center}
   \caption[example] 
   { \label{fig:same_r0} 
Heatmap of EOF predictor performance (in median RMS error in nm) as training data length \textbf{l} and history vector length \textbf{n} vary. Left: windspeed of 7 m/s. Middle: windspeed of 11 m/s. Right: windspeed of 14 m/s. All of the above simulations have an r$_0$ of 15 cm. Note that the optimal history vector length is different for the slowest atmosphere. Given the two right-most plots find an optimal in the corner of our grid, they may also find different optimal values given an expanded search space.}
   \end{figure}

For each hyper-parameter grid search, we build a predictive EOF filter with varying numbers of frames of training data (\textbf{l}) and varying history vector lengths (\textbf{n}), and build a heat map of the the median performance in root mean square (RMS) error. We compute a residual by subtracting the prediction of the full phase-screen at each iteration from the known injected phase at each iteration, and take an RMS of that difference at each point in time, leaving us with the RMS for each of 10000 iterations. We then take the median of that time-series distribution for the final value plotted and used for optimization. Figures \ref{fig:same_speed} and \ref{fig:same_r0} show examples of differing r$_0$ with the same windspeed and differing windspeed with the same r$_0$ respectively. We note that the optimal history vector length changes based on different turbulence conditions. 


While these simulations show that the optimal history vector length changes, simulations  consistently show that the optimal amount of training data is the maximum amount available. We expect that predictor performance due to different lengths of training data in the filter matrix would be more impactful on real data, where including more frames of data introduces more noise into the training (our simulations are lacking measurement noise), and when the velocity layer will vary over some timescale.

\section{ON-THE-FLY OPTIMIZATION WITH SIMULATED ANNEALING}

If the desired strategy for EOF implementation is to do on-sky hyper-parameter tuning, building a full grid of filter hyper-parameter runs is likely not feasible. For these initial grids, we used 10 seconds of data and built a filter based on up to 24 seconds of training data; even for a system optimized to run in real time, to do this 600 times would take $\sim$ 6 hours. We present an alternative way to explore this hyper-parameter space more optimally with simulated annealing, a probabilistic method for finding the global optimum of a non-differentiable function. 

  \begin{figure} [h!]
   \begin{center}
   \begin{tabular}{c} 
   \includegraphics[width=0.45\linewidth]{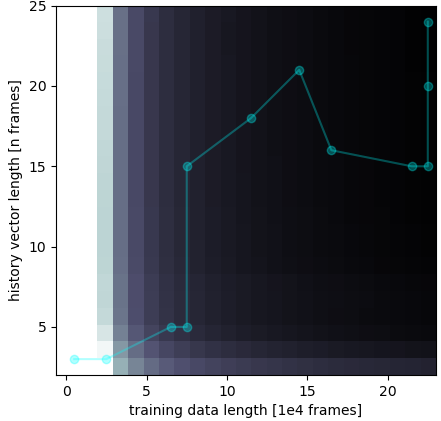}
\includegraphics[width=0.45\linewidth]{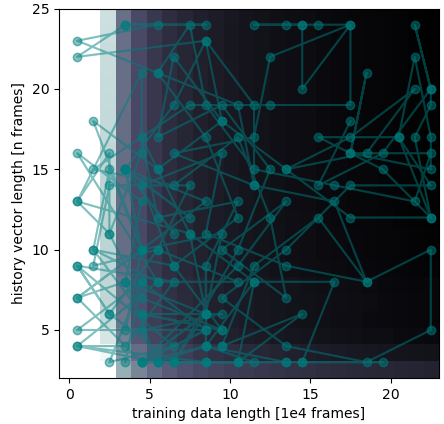}
   \end{tabular}
   \end{center}
   \caption[example] 
   { \label{fig:sim_anneal_demo} 
Simulated annealing samples many values, but it does not accept every step. This is the run of a simulated annealing algorithm applied to a grid with a windspeed of 14 m/s and an r$_0$ of 15 cm, chosen because the grid had the most variation. Left: Entire walk of the simulated annealing to find the optimal solution. Right: Every point sampled (whether or not it was accepted) over the 400 steps.}
   \end{figure} 

Simulated annealing \cite{stats_textbook} is a probabilistic way to explore a discrete parameter space. The algorithm picks a random point of a given step size away from the current point and accepts it if the point provides a better solution, or if a Metropolis criterion is met with a given temperature. This process is named after metal annealing, where metal is heated and cooled strategically for optimal performance. Pseudo-code of this algorithm and more discussion of the Metropolis criterion are presented in Appendix A. 
Figure \ref{fig:sim_anneal_demo} shows how the algorithm walks through one of the hyper-parameter search grids shown above. 

   \begin{figure} [ht]
   \begin{center}
   \begin{tabular}{c} 
   \includegraphics[width=0.48\linewidth]{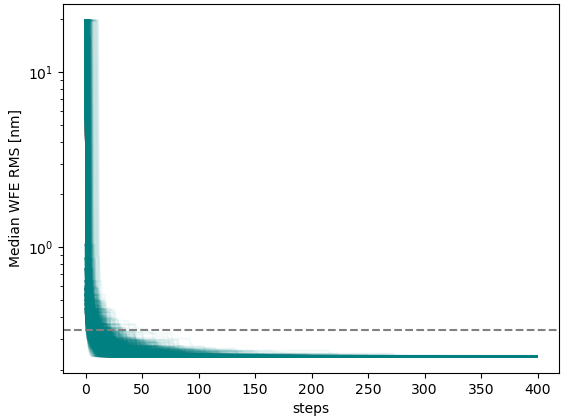}
\includegraphics[width=0.48\linewidth]{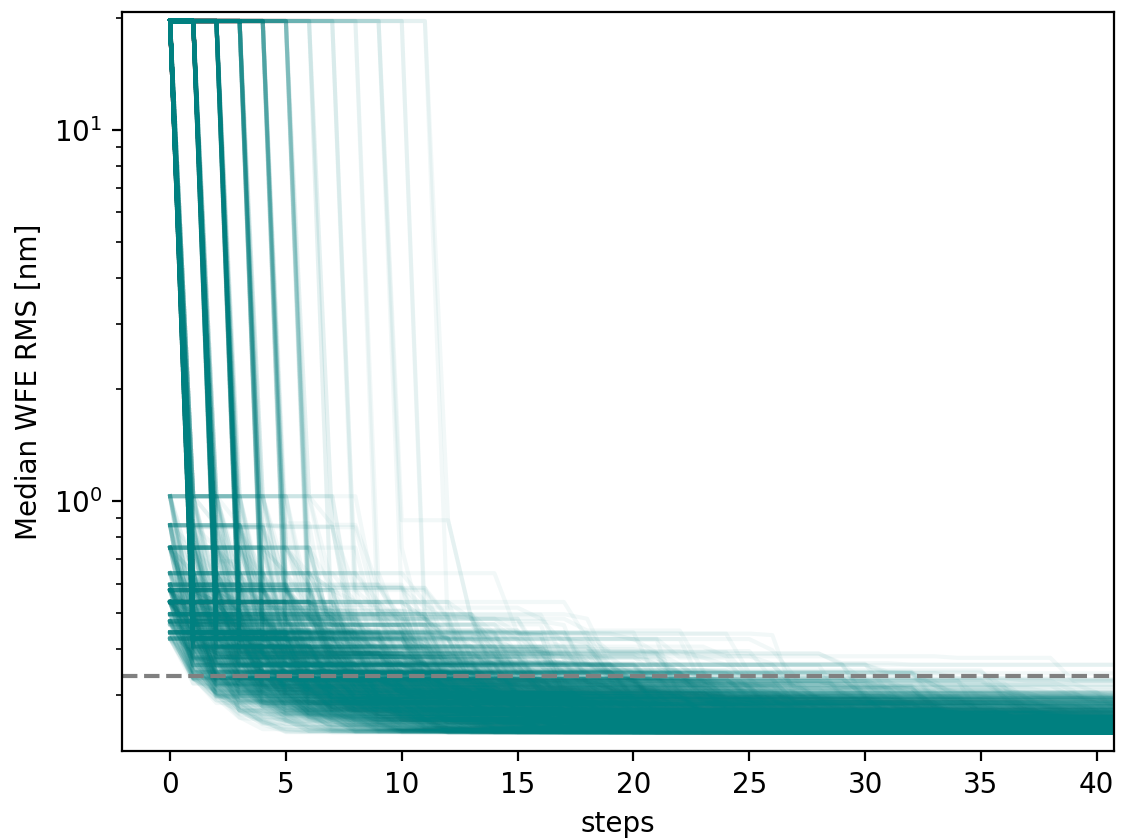}
   \end{tabular}
   \end{center}
   \caption[example] 
   { \label{fig:annealing_performance} 
Simulated annealing is applied to the previously shown dataset, with a windspeed of 14 m/s and an r$_0$ of 15 cm. The input temperature is 119 and the step size is 5. The dashed line indicates when the algorithm reaches within $0.1$ nm of the optimal solution. Left: full run. Right: closer detail on the first few iterations.}
   \end{figure}

We applied simulated annealing to the previously shown filter hyper-parameter grid for a windspeed of 14 m/s and an r$_0$ of 15 cm, and found that the algorithm could repeatably find within $0.1$ nm of the optimal value within 10 steps and repeatably recover the exact solution within 400 steps. For an on-sky strategy, we expect we could use closer to 10 steps, to find a very good if not exactly optimal solution. However, how quickly the simulated annealing algorithm converges to near its minimum will be dependent on the performance of each filter over a set of conditions, so testing with lab-data and on-sky telemetry will be needed before specific strategies can be recommended. 

Figure \ref{fig:annealing_performance} shows 1000 runs of simulated annealing on the same data set. This example was run with a step size of 5 (which maps to a single history vector frame and 50000 training data frames), and a temperature of 119. Initial testing found that higher temperatures (100-200) worked better for this data set but no quantitative optimization was done for temperature and step size in this case, due to how well the algorithm performed with these parameters. A strength of simulated annealing is its effective application to problems without significant optimization or previous knowledge of a parameter space.

\section{PRELIMINARY LAB RESULTS}

To further this work we plan to examine $\alpha$ as third hyper-parameter to optimize, as well as explore how more realistic conditions (e.g., imperfect noisy data) impact the optimal training data length. To do this, we plan to run similar grids of optimization on the SEAL (Santa Cruz Extreme AO Laboratory) testbed \cite{Jensen-Clem2021}. Along with more realistic conditions, SEAL has a coronagraphic branch with a vector vortex coronagraph \cite{Moreno2024}; moving to the bench will allow us to optimize hyper-parameters based on coronagraphic contrast as well as wavefront error. 

\begin{figure}[h!]
\makebox[\textwidth][c]{
\mbox{\includegraphics[width=0.8\textwidth]{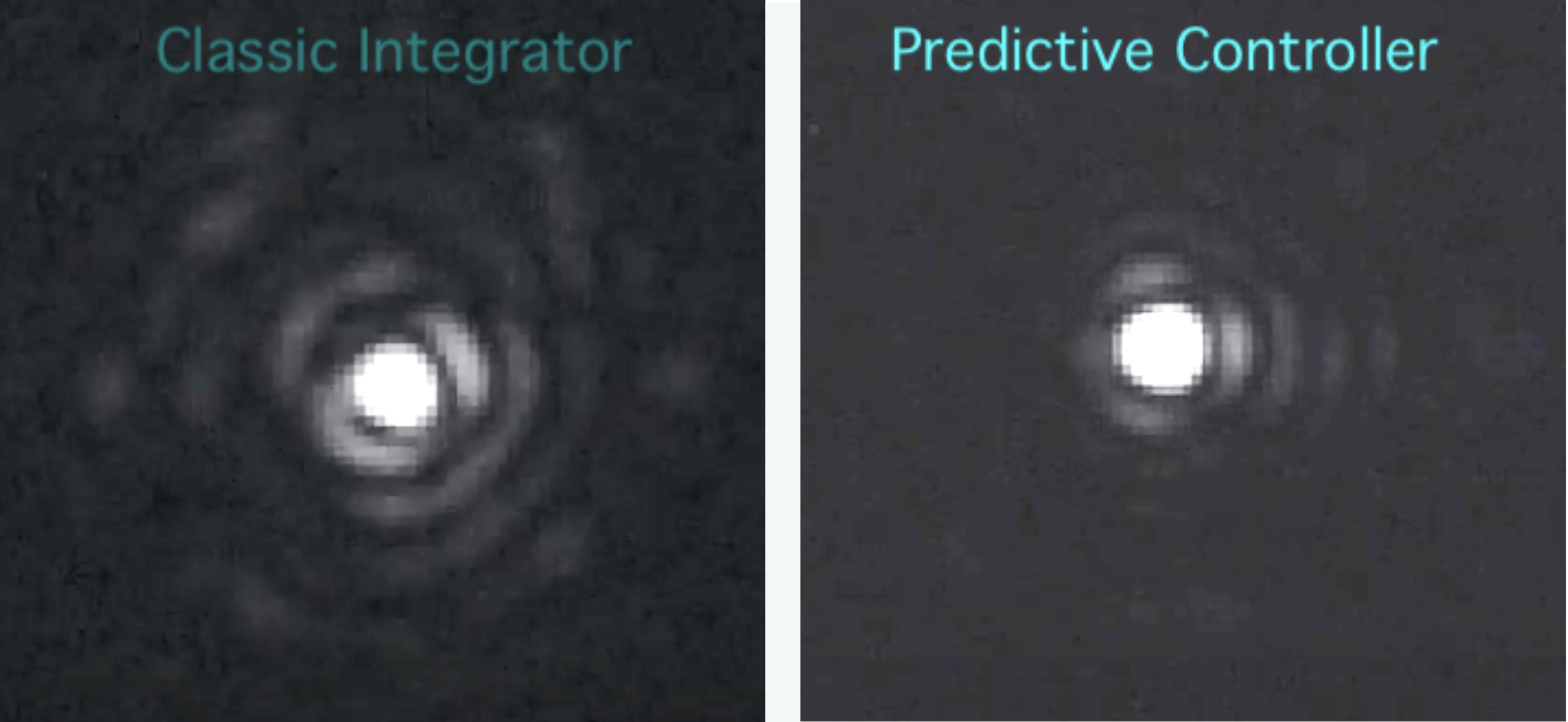}}
}
    \caption{PSFs of predictive control and a classic integrator side-by-side on the same color-scale. Note: these are \textit{preliminary} results without robust image calibration and serve only as an initial visual comparison. }
\label{fig:psf}
\end{figure}

\begin{figure}[h!]
\makebox[\textwidth][c]{
\mbox{\includegraphics[width=0.6\textwidth]{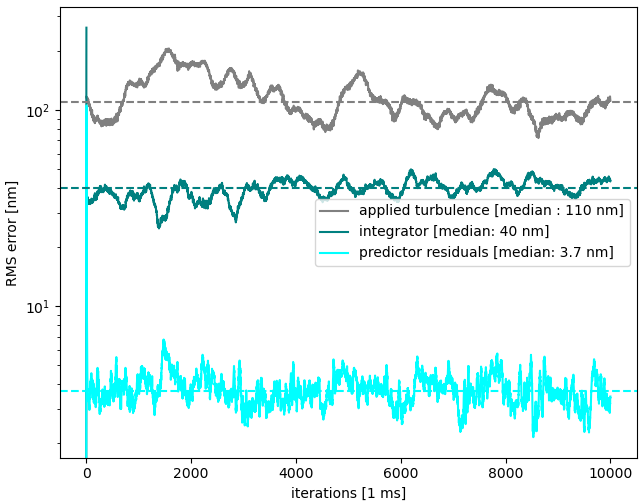}}
}
\caption{Bench results of an EOF controller as compared to a classic integrator. RMS wavefront is estimated from the SHWFS residuals as converted to DM space. Input turbulence is a single wind layer with a pupil crossing time of 1s and an atmosphere with a D/r$_0$ of 100, with amplitude scaled down the linear $2\pi$ dynamic range of the Boston MEMS. 
}
\label{fig:wfe}
\end{figure}

Figures \ref{fig:psf} and \ref{fig:wfe} show a preliminary implementation of EOF predictive control on the SEAL testbed, comparing the performance of EOF (with a filter built with a 5 frame history vector and 60000 frames of training data) to an integrator with a gain of 0.2. We simulate a single layer atmosphere with a D/r$_0$ of 100 (e.g., a 10 meter telescope with an r$_0$ of 10 cm) and a pupil crossing time of 1s (e.g., 10 m/s wind layer) and sample its phase every millisecond. We apply turbulence with the same D/r$_0$ in the lab, but with the amplitude scaled down to the linear regime of the Boston MEMS DM (within 2$\pi$ of phase), and then applied and corrected with the MEMS DM for 10000 iterations.

We use a high speed Thorlabs Shack-Hartmann Wavefront Sensor with 640 slopes and a Boston Micromachines MEMS deformable mirror with 24 actuators across the pupil. Initial results show improvement over a classic integrator both in reduction of wavefront error (Figure \ref{fig:wfe}), and in the the focal plane PSF images (Figure \ref{fig:psf}). 

Plans for future lab testing include recreating these hyper-parameter grid searches in an environment more indicative of a true AO system; we will apply turbulence with a Meadowlark spatial light modulator of 1116 pixels across, and correct with a woofer/tweeter system of a 97 actuator ALPAO deformable mirror and the aforementioned Boston MEMS. This will give us the ability to mimic more realistic effects of measurement error (with turbulence applied as a much finer spatial scale than we can correct.) 
   

\section{CONCLUSIONS}
\label{sec:misc}

In conclusion, we explore future facilitization of predictive wavefront control by looking at optimal hyper-parameters for the training data and history vector length of the predictive filter. We find that in simulation, optimal filter hyper-parameters vary based on the atmospheric turbulence, and expect that we will need to optimize on-sky. We explore simulated annealing as a way to probe filter hyper-parameters more efficiently, and find it can repeatably converge to within 0.1 nm RMS of the optimal solution within 10 iterations. 

We present preliminary lab results, that compare EOF to a classic integrator, and see promising performance from EOF. This lab implementation is a precursor to future exploration of $\alpha$ in the EOF matrix inversion, as well as an opportunity to study the impact of these corrections on the coronagraphic dark hole. We also hope to test this optimization on-sky with Keck/NIRC2 or Subaru/SCExAO. 

Facilitizing predictive control for regular on-sky use will not only improve current high contrast imaging for large telescopes, but will show technology maturation for its use in upcoming extremely large telescopes. When predictive control is in use, the future of our coronagraphic dark hole is looking dark. 

\appendix    

\section{SIMULATED ANNEALING ALGORITHM}
\label{sec:simulated_annealing}

Here we further describe the Metropolis criterion, and present \texttt{Python} pseudo-code of the simulated annealing algorithm to demonstrate the application of this algorithm to our data. For each iteration, the algorithm picks a new point, evaluates if that point is a better solution, and if not decides to keep the new point only if it meets the Metropolis criterion, which is met by comparing a randomly generated number to a function built from a cooling input temperature and how far the evaluation of the new point is from the previous. We note that simulated annealing and the Metropolis criterion make this algorithm vulnerable to getting trapped in a deep local minimum, and plan to study this impact with more realistic lab data. For a given iteration i: 
\begin{align}
    &\textrm{cooled temperature} = \frac{\textrm{temperature}}{i + 1} \\
    &\textrm{difference} = \textrm{evaluation}(i- 1) - \textrm{evaluation}(i) \\
    &\textrm{metropolis} = \exp\left(\frac{\textrm{difference}}{\textrm{cooled temperature}}\right)
\end{align}

This logic runs for a given number of steps, but on-sky this could also be a cutoff based on how much the algorithm is still improving the solution or some minimum allowable wavefront error. 

\begin{figure}[h!]
\centering
\begin{lstlisting}
def simulated_annealing(objective, bounds, n_iterations, step_size, temp): 
    # start with an initial guess
    best = preset or random guess
    # evaluate it and set it as a starting point
    best_eval = objective(best)
    current, current_eval = best, best_eval 
    # for every iteration
    for i in range(n_iterations): 
        # take a step 
        candidate = move from current 
        # evaluate the new candidate
        candidate_eval = objective(candidate) 
        # check for new best solution 
        if candidate_eval < best_eval: 
            # store new best point 
            best, best_eval = candidate, candidate_eval
        # otherwise, use metropolis to see if we keep the new point 
        # or stay with the previous guess 
        # this determines where the guess next iteration starts from
        diff = candidate_eval - current_eval 
        t = temp/(i + 1)  
        # calculate metropolis acceptance criterion 
        metropolis = np.exp(-diff / t)  
        
        # check if we should keep the new point 
        if diff < 0 or random number < metropolis: 
            # store the new current point 
            current, current_eval = candidate, candidate_eval 

    # after however many iterations return the best solution
    return best, best_eval
\end{lstlisting}
\caption{Pseduo-code of simulated annealing algorithm, heavily influenced by \href{https://machinelearningmastery.com/simulated-annealing-from-scratch-in-python/}{Machine Learning Mastery}. The details of making consistent integer steps over a two dimensional grid make this algorithm more complex, so we have chosen to present representative code as opposed to the literal implementation.}
\label{fig:weatherCal}
\end{figure}

\newpage
\acknowledgments 

This work was supported by NSF ATI Grant 2008822. Thank you to the folks at Keck Observatory and members of the SCExAO team for discussing their implementations of EOF.

\bibliography{report} 
\bibliographystyle{spiebib} 

\end{document}